\documentclass[11pt,notitlepage,preprintnumbers,longbibliography,nofootinbib]{article}
\pdfoutput=1 
\usepackage{jheppub}
\usepackage[utf8]{inputenc}
\usepackage[utf8]{inputenc}
\usepackage{amsmath}
\usepackage{amssymb}
\usepackage{commath}
\usepackage[mathscr]{euscript}
\usepackage{color}
\usepackage{physics}
\usepackage{braket}
\usepackage{tensor}
\usepackage{float}
\usepackage{soul}
\usepackage{hyperref}
\usepackage{multirow}
\usepackage[dvipsnames]{xcolor}
\usepackage{bclogo}
\usepackage{comment}
\usepackage{dsfont}
\usepackage{cancel}
\usepackage[normalem]{ulem} %sout
\usepackage[T1]{fontenc}
\usepackage{pifont}
\usepackage{bm}
\renewcommand{\dd}{\mathrm{d}}

\DeclareMathOperator{\arcsinh}{arcsinh}
\DeclareMathOperator{\arctanh}{arctanh}

\DeclareMathOperator{\arccoth}{arccoth} 
\makeatletter
\def\@fpheader{\relax}
\makeatother

\begin{document}
\title{Temporal Entanglement Entropy as a probe of Renormalization Group Flow}
\author[a]{Sebastian Grieninger,}
\author[a,b]{Kazuki Ikeda,}
\author[a,b,c]{Dmitri E. Kharzeev}
\affiliation[a]{Center for Nuclear Theory, Department of Physics and Astronomy,
Stony Brook University, Stony Brook, New York 11794–3800, USA}
\affiliation[b]{Co-design Center for Quantum Advantage (C2QA), Stony Brook University, USA}
\affiliation[c]{Department of Physics, Brookhaven National Laboratory
Upton, New York 11973-5000, USA}
\emailAdd{sebastian.grieninger@stonybrook.edu}
\emailAdd{kazuki.ikeda@stonybrook.edu}
\emailAdd{dmitri.kharzeev@stonybrook.edu}
\date{\today}
\abstract{
The recently introduced concept of timelike entanglement entropy has sparked a lot of interest. Unlike the traditional spacelike entanglement entropy, timelike entanglement entropy involves tracing over a timelike subsystem. In this work, we propose an extension of timelike entanglement entropy to Euclidean space (``temporal entanglement entropy"), and relate it to the renormalization group (RG) flow. Specifically, we show that tracing over a period of Euclidean time corresponds to coarse-graining the system and can be connected to momentum space entanglement. We employ Holography, a framework naturally embedding RG flow, to illustrate our proposal. Within cutoff holography, we establish a direct link between the UV cutoff and the smallest resolvable time interval within the effective theory through the irrelevant $T\bar T$ deformation. Increasing the UV cutoff results in an enhanced capability to resolve finer time intervals, while reducing it has the opposite effect. Moreover, we show that tracing over a larger Euclidean time interval is formally equivalent to integrating out more UV degrees of freedom (or lowering the temperature). As an application, we point out that the temporal entanglement entropy can detect the critical Lifshitz exponent $z$ in non-relativistic theories which is not accessible from spatial entanglement at zero temperature and density.}
\maketitle
\section{Introduction}
Recently, the authors of \cite{Doi:2023zaf} introduced a new complex-valued measure of information -- timelike entanglement entropy (EE) -- which attracted a lot of interest~\cite{Doi:2022iyj,Reddy:2022zgu,Diaz:2021snw,Li:2022tsv,Liu:2022ugc,Narayan:2022afv,Alshal:2023kcd,Foligno:2023dih,Chen:2023prz,Chen:2023gnh,Jiang:2023ffu,Narayan:2023ebn,Jiang:2023loq,Chu:2023zah,He:2023wko,Franken:2023pni,Chen:2023sry,Kawamoto:2023nki,Chen:2023eic,Parzygnat:2023avh,He:2023ubi,Carignano:2023xbz,Guo:2023aio,Aguilar-Gutierrez:2023tic,Omidi:2023env,Narayan:2023zen,Shinmyo:2023eci,Guo:2023tjv,Apolo:2023ckr,Kanda:2023jyi,He:2023syy}. To compute a timelike EE, one traces over a timelike subsystem instead of a space-like one. In this work, we suggest to extend the timelike entanglement entropy to Euclidean space, and interpret the resulting ``temporal EE" in terms of the renormalization group (RG) flow. In particular, tracing over Euclidean time corresponds to coarse-graining the system and is thus related to momentum space entanglement entropy which was introduced and studied in \cite{Balasubramanian:2011wt,Costa:2022bvs}. In holography, the concept of RG flow is naturally embedded since the additional bulk dimension is related to the energy scale of the boundary quantum field theory \cite{deBoer:1999tgo,deBoer:2000cz,Fukuma:2002sb,Skenderis:2002wp}. Moreover, it is straightforward to compute entanglement entropy in holography using the Ryu-Takayanagi formula \cite{Ryu:2006ef,Ryu:2006bv,Hubeny:2007xt}. Therefore, holography is the ideal framework to illustrate our proposal.

\begin{figure}
    \centering
    \includegraphics[width=0.7\linewidth]{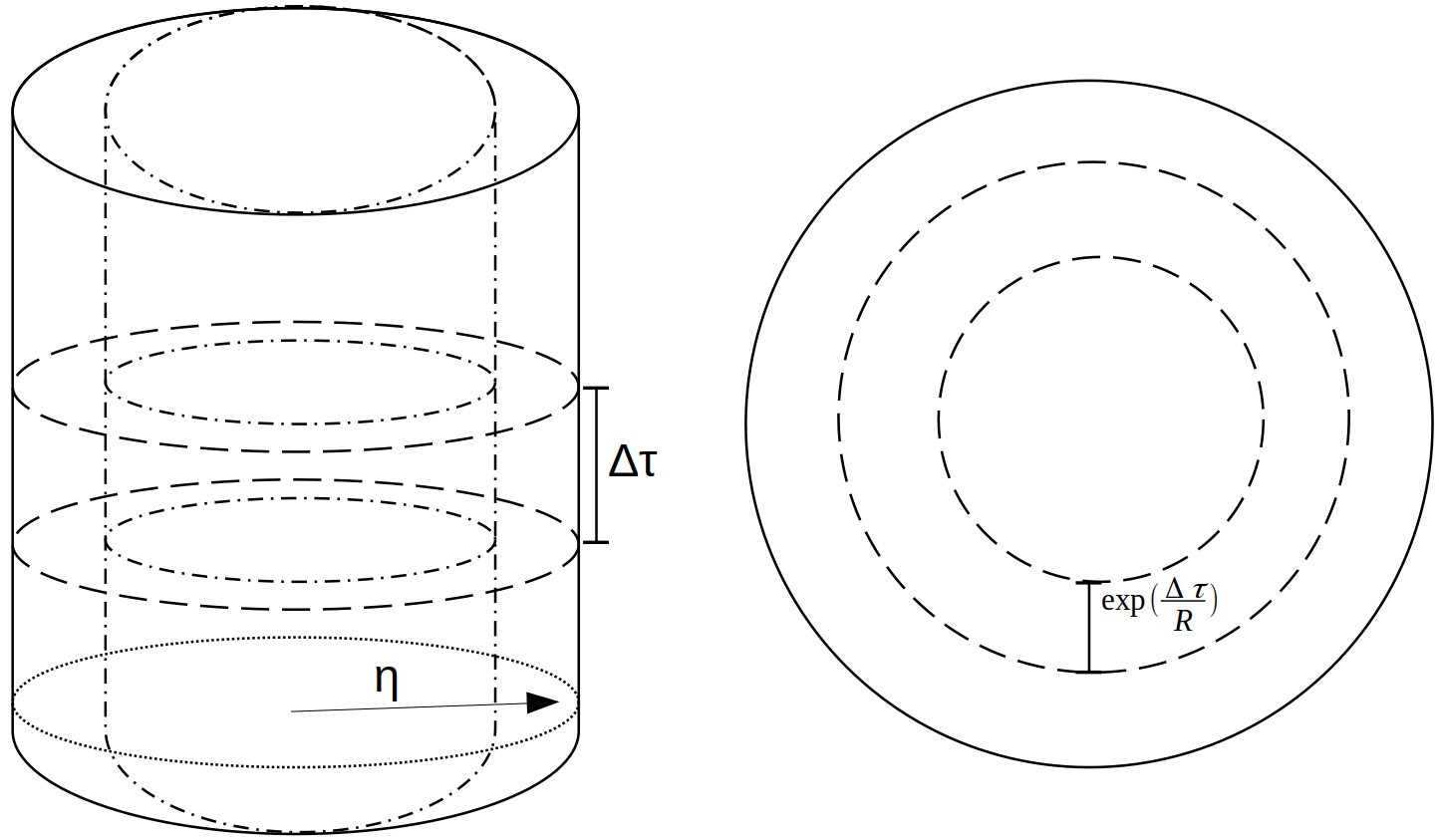}
    \caption{Relation of AdS cylinder in global coordinates and CFT in
radial quantization. Time translations in the bulk correspond to dilations in CFT linking energies in AdS to dimensions in CFT. Figure adapted from \cite{KaplanJHU}. The long dash dotted lines in the left plot correspond to the cutoff geometry which is dual to a $T\bar T$ deformation of the field theory.}
    \label{fig:1}
\end{figure}

Consider global Euclidean AdS$_3$ in global coordinates \begin{equation}
    \dd s^2=L^2(\dd \eta^2+\cosh(\eta)^2\,\dd \tau^2+\sinh(\eta)^2\,\dd\phi^2).
\end{equation}
We can introduce a new radial coordinate $\theta\in[0,\pi/2]$ by $\tan(\theta)=\sinh(\eta)$
\begin{equation}
    \dd s^2=\frac{L^2}{\cos(\theta)^2}\,(\dd\eta^2+\dd\tau^2+\sin(\theta)^2\,\dd\phi^2).
\end{equation}
Since we restrict $\theta\in[0,\pi/2]$ the metric covers only half of $\mathbb{R}\times S^2$. If we approach the conformal boundary as $\eta=\pi/2-\varepsilon\,e^{-\tau}$ with $\epsilon\to 0$, we find the effective boundary metric~\cite{KaplanJHU}
\begin{equation}
    \dd s^2_{\partial AdS_3}\to e^{2\tau}(\dd \tau^2+\dd\phi^2)\to \dd \rho^2+\rho^2\dd \phi^2.
\end{equation}
In the last step we introduced $\rho=e^\tau$. The origin of the Euclidean plane, $\rho=0$, maps to the infinite past of Euclidean AdS at $\tau=-\infty$ and $\rho=\infty$ describes the infinite future. Figure \ref{fig:1} illustrates this mapping relating the AdS cylinder in global coordinates and the CFT in radial quantization. In this context, the time translation operator in the AdS bulk corresponds to the dilatation operator in the CFT, linking energies in AdS to dimensions in the CFT. 

This is particularly interesting in view of the surface/state correspondence~\cite{Miyaji:2015fia,Miyaji:2015yva,Caputa:2017urj}. As outlined in \cite{Goto:2017olq}, we can describe excitations in the field theory at $(t=0,\phi)$ by inserting primary operators at $(\tau_0,\phi), (-\tau_0,\phi)$. The bulk position $\eta$ is determined by considering the intersection of the geodesic (in Euclidean time) with the plane at $\tau=0$.

Furthermore, we can expand upon this concept to encompass states $\psi_{O(\tau)}$ that result from inserting an operator $O(\tau)$ at $\tau \to - \infty$ ~\cite{KaplanJHU}. Placing an operator at the origin ($\rho = 0$, or $\tau \to - \infty$)  within the CFT delineates a particular AdS/CFT state. When interpreted within holography, this configuration establishes an initial state in the remote past, evolving into $\psi_{O(\tau)}$ at a finite moment in Euclidean time. It is important to recognize that $e^\tau \sim |\rho|$, implying that the global coordinate time corresponds to the logarithm of the radius of a circle centered at the origin within the CFT.

Both examples illustrate the deep connection between Euclidean time and RG flow for conformal theories. In this manuscript, we suggest that the concept of temporal entanglement entropy corresponds  to coarse-graining in field theory. If we trace over late Euclidean times, we lose fine-resolution information about the state.  

In this work, we present a geometric version of the RG flow triggered by the $T\bar T$ operator ($T$ is the energy-momentum tensor) in two-dimensional conformal field theories at large central charge. The $T\bar T$ deformation has received a lot interest since it is an exactly  solvable irrelevant deformation that is well-defined up to arbitrary scales~\cite{Smirnov:2016lqw,Cavaglia:2016oda,Zamolodchikov:2004ce}. We utilize cutoff holography \cite{McGough:2016lol} which is the holographic realization of the $T\bar T$ deformation (in the absence of matter) to derive a connection between the smallest (Euclidean) time interval that we can resolve and the RG scale. The $T\bar T$ deformation is \textit{moving the boundary theory into the bulk}. The outcome is that the effective theory resulting from the deformation will possess a cutoff scale which roughly corresponds to the cutoff radius within the bulk~\cite{Guica:2019nzm}.

%\section{Timelike Entanglement Entropy}
In the context of quantum field theory, the concept of timelike entanglement entropy can be related to the usual notion of spacelike entanglement entropy by a Wick rotation, that changes a spacelike boundary subregion into a timelike one. The authors of \cite{Doi:2023zaf} established a precise definition of timelike entanglement entropy in 2d field theories and demonstrated agreement with the a computation via the replica trick.

The authors posit that timelike entanglement entropy should be correctly interpreted as ``pseudo entropy'' \cite{Doi:2022iyj}, which corresponds to the von Neumann entropy of a reduced transition matrix. In the context of holographic systems, they define timelike entanglement entropy as the total complex-valued area of a specific stationary combination of both spacelike and timelike extremal surfaces, provided these surfaces are homologous to the boundary region.

\section{$T\bar T$ deformations}

Irrelevant deformations within Quantum Field Theories (QFTs) are still poorly understood partly because the UV fixed point may not be well-defined. %because one cannot follow the standard renormalization group procedure, as we now explain. An Infrared (IR) -relevant coupling is irrelevant in the Ultraviolet (UV). However, for IR-irrelevant operators (which are UV relevant), we usually cannot start from the UV as desired from an IR effective theory perspective. 
For this particular reason, the %IR-
irrelevant $T\bar T$ deformation gained a lot of attention~\cite{Smirnov:2016lqw,Cavaglia:2016oda,Zamolodchikov:2004ce}: unlike generic irrelevant deformations, the $T\bar T$ deformation is exactly solvable.

\subsection{\label{sec:2d}$T\bar T$ Deformations in $2d$ Field Theory}

In this subsection, we give a brief summary of $T\bar T$ deformations in 2 dimensional field theory. For a more detailed introduction we refer the reader to \cite{Jiang:2019epa,GuicaCERN} and references therein. 

We can deform a seed conformal field theory by the irrelevant deformation $\det(T_{\mu\nu}^{(t)})$:

\begin{equation}
    S_\text{QFT}=S_\text{CFT}+2\pi\mu\,\int\dd^2x\,\sqrt{\gamma}\,T\bar T,\label{eq:deformac}
\end{equation}
where the resulting stress tensor of the deformed theory is no longer traceless and $T\bar T=\frac18\left(T^{\mu\nu}T_{\mu\nu}-(T_\mu^\mu)^2\right)$ in $d=2$; $\gamma$ is the metric. The composite $T\bar T$ operator has dimension 4.
If we have a single mass scale $\lambda$ in the theory, then we can re-write the deformation on the level of the action, i.e. $\dd S/\dd\mu=\int \dd^2x\,T\bar T$ as defined in eq. \eqref{eq:deformac} (using $\lambda=1/\sqrt{\mu}$)
\begin{equation}
    \lambda\frac{\dd S_\text{QFT}}{\dd\lambda}=\int\dd^2x\,\sqrt{\gamma}\, T^\mu_\mu.\label{eq:traceflow}
\end{equation}
Eq. \eqref{eq:traceflow} shows that we can re-cast the $T\bar T$ deformation in form of a trace-flow equation which \cite{McGough:2016lol} used to write down an expression for the $T\bar T$ deformation in the language of holography.
\subsection{$T\bar T$ Deformations in Holography}
McGough, Mezei, and Verlinde~\cite{McGough:2016lol} proposed that the holographic dual of the $T\bar T$ deformation corresponds to cutting off the space of the gravity theory at a finite radial position. If we consider AdS$_3$ in a radial slicing like
\begin{align}
    \dd s^2=\dd \eta^2+g_{\mu\nu}(\eta,x)\,\dd x^\mu\,\dd x^\nu, && \text{with } \eta<\eta_c,\label{eq:AdSsub}
\end{align}
the condition $\eta<\eta_c$ implements this restriction~\cite{Hartman:2018tkw,Kraus:2018xrn, Gorbenko:2018oov,Grieninger:2019zts,Grieninger:2020wsb}. 

Due to the finite radial cutoff, the Conformal Field Theory (CFT) is no longer located at $\eta=\infty$ as usual but rather at a finite radial distance $\eta=\eta_c$. Depending on the slicing of AdS, the field theory may thus live on a curved space instead of the flat boundary. The central charge $c$ is related to the deformation parameter $\mu$ by~\cite{Brown:1986nw}:

\begin{equation}\label{eq:TTbarcentralcharge}
c=\frac{3L}{2G_N},\ \ \mu=16\pi G_N L,
\end{equation}
where $L$ is the curvature radius of the AdS space, and $G_N$ is the Newton constant of the gravity theory. 
In this work, we assume that the central charge $c$ is large and use the weak form of the AdS/CFT correspondence.

The renormalized Einstein-Hilbert action in AdS$_3$ is given by
\begin{equation}
    S=-\frac{1}{16\pi\,G_N}\int_{\mathcal M}\!\!\dd^3x\,\sqrt{g}\left(R-\frac{2}{L^2}\right)-\frac{1}{8\pi\,G_N}\int_{\partial M}\!\!\dd^2x\,\sqrt{\gamma}\left(K-\frac 1L\right),
\end{equation}
where $g_{mn}$ is the 3-dimensional metric of AdS, $\gamma_{\mu\nu}$ is the 2-dimensional metric on the cutoff slice and $K$ is extrinsic curvature. 
The corresponding renormalized holographic stress tensor reads~\cite{deHaro:2000vlm}
\begin{equation}
    T_{\mu\nu}=\frac{1}{8\pi\,G_N}\left(K_{\mu\nu}-K\,\gamma_{\mu\nu}+\frac{\gamma_{\mu\nu}}{L}\right).\label{eq:em}
\end{equation}
 We can derive the flow equation by considering the trace of the radial Einstein equation which in $d=2$ is given by
\begin{equation}\label{eq:radE}
    G^\eta_\eta=\frac 12\,\left(K^2-K^{\mu\nu}K_{\mu\nu}\right)-\frac{1}{L^2}-\tilde R=0,
\end{equation}
where we denote the Ricci scalar on the curved cutoff slice $\eta=\eta_c$ by $\tilde R$. We can obtain an explicit expression for the $T\bar T$ operator by evaluating $T\bar T=(T^{\mu\nu}T_{\mu\nu}-(T_\mu^\mu)^2)$ for the energy-momentum tensor \eqref{eq:em}. We can eliminate the extrinsic curvature $K$ from the expression by solving the radial Einstein equation \eqref{eq:radE} for $K$ and plugging it into the energy-momentum tensor \eqref{eq:em}. With this, the holographic expression for the trace flow equation of the deformed energy-momentum tensor simplifies to
\begin{equation}
    T_\mu^\mu=-\frac{L}{16\pi G_N}\tilde R-4\pi\,G_NL\,(T^{\mu\nu}T_{\mu\nu}-(T_\mu^\mu)^2)=-\frac{c}{24\pi}\tilde R-\frac{\mu}{4}\,T\bar T.
\end{equation}
For the case of finite temperature, see for example \cite{Chen:2018eqk}. 

In general, the $T\bar T$ deformation should be implemented by a double trace deformation on the original conformal boundary (since the theory is UV complete) \cite{Guica:2019nzm}. However, in \cite{Guica:2019nzm} it was shown that in the absence of matter and for the specific sign of the $T\bar T$ deformation that we are using the double trace deformation is equivalent to considering cutoff holography.

\section{Temporal Entanglement Entropy and RG flow}

In this work, we restrict ourselves to $d=2+1$ dimensions of the gravity theory corresponding to two-dimensional field theory. The generalization to higher dimensions is straightforward. We consider global AdS$_3$ 
\begin{equation}
    \dd s^2=L^2(\dd \eta^2+\cosh(\eta)^2(-\cosh(r)^2\dd t^2+\dd r^2))=L^2(\dd \eta^2-\cosh(\eta)^2\,\dd t^2+\sinh(\eta)^2\,\dd\phi^2)
\end{equation}
at a fixed spatial position $\phi=0$. In this coordinate system the conformal boundary is located at $\eta=\infty$. For spacelike geodesics the induced metric on the worldline is given by
\begin{equation}
    \dd s_\text{ind}^2=L^2(1-\cosh(\eta)^2\,t'(\eta)^2)\,\dd\eta^2=L^2(1+\cosh(\eta)^2\,\tau'(\eta)^2)\,\dd\eta^2,
\end{equation}
where we performed a Wick rotation to Euclidean time $\tau=it$ in the last step. In Euclidean signature, we have to minimize the area functional 
\begin{equation}
    A=L\,\int\dd \eta\,\sqrt{1+\cosh(\eta)^2\,\tau'(\eta)^2}.\label{eq:area}
\end{equation}
The associated equations of motion are given by
\begin{equation}
   L \frac{\tau'(\eta)\cosh(\eta)^2}{\sqrt{1+\cosh(\eta)^2 \tau'(\eta)^2}}=c_1
\end{equation}
with solution 
\begin{equation}
    \tau(\eta)=c_2\pm\arctanh\left(\frac{c_1/L \sinh(\eta)}{\sqrt{1+c_1^2/L^2+\sinh(\eta)^2}}\right).\label{eq:sol}
\end{equation}
The two constants are determined by imposing that the geodesic vanishes at the turning point $\eta_\star$ with infinite derivative, i.e. $\tau(\eta_\star)=0$ and $\tau'(\eta_\star)=\infty$.
Imposing the two boundary conditions, we find $c_2=\mp i\pi/2$ and $c_1=\cosh(\eta_\star)$ and the solution reads
\begin{equation}\label{eq:minimal}
    \tau(\eta)=\pm\arctanh\left(\frac{\cosh(\eta_\star)\sinh(\eta)}{\sqrt{\cosh(\eta)^2-\cosh(\eta_\star)^2}}\right)\pm\frac{i\pi}{2}.
\end{equation} 
 Note that the entangling surfaces are real since the argument of the hyperbolic arctangent is larger than 1 and we could use the identity  $\arctanh(x)=\pm \pi i/2+1/2\log((x+1)/(x-1))$ (valid for $x\in(-\infty,-1)\cup (1,\infty)$) to re-write the expression.

\begin{figure}
    \centering
    \includegraphics[width=0.6\linewidth]{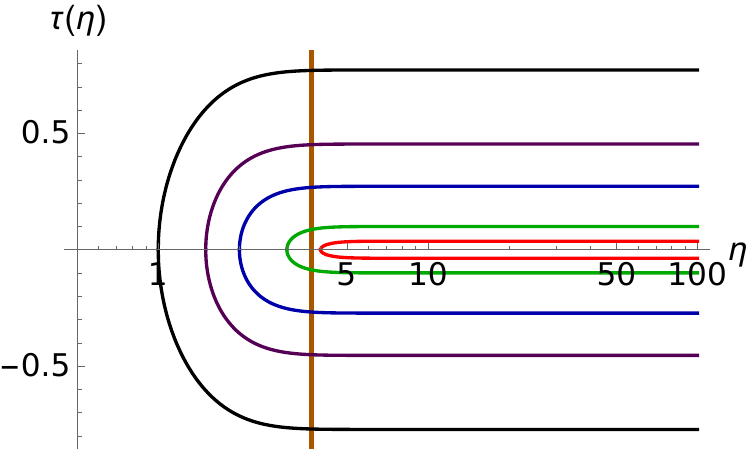}
    \caption{Minimal surfaces \eqref{eq:minimal} as a function of the radial coordinate $\eta$ for different turning points in the bulk. To probe larger time intervals $T_0$ we require information from deeper in the bulk (the IR), i.e. the turning point moves so smaller $\eta$. Moving the cutoff into the bulk results into losing the minimal surfaces that resolved the smallest time intervals in the original theory.}
    \label{fig:entanglingsurfaces}
\end{figure}

The turning point is related to the length of the time interval at the boundary by $T_0=\tau_+(\infty)-\tau_-(\infty)=2 \log(\coth(\eta_\star/2))$. We illustrate the solutions \eqref{eq:minimal} for various time intervals $T_0$ in figure \ref{fig:entanglingsurfaces}.

Evaluating the area functional \eqref{eq:area} on the solution \eqref{eq:sol}, we find
\begin{align}
    A&=2L\int_0^{\eta_c}\dd\eta\sqrt{\frac{\cosh (2 \eta)+1}{\cosh (2 \eta)-\cosh (2 \eta_\star)}}=2L\arctanh\left(\frac{\sqrt{2} \sinh (\eta_c)}{\sqrt{\cosh (2 \eta_c)-\cosh (2 \eta_\star)}}\right)+i\pi\,L.\label{eq:area1}
\end{align}
The area is real since the argument of the inverse hyperbolic tangent is larger than 1. We can make this more explicit by using that the inverse hyperbolic tangent satisfies $\arctanh(x)=\pm \pi i/2+1/2\log((x+1)/(x-1))$ for $x\in(-\infty,-1)\cup (1,\infty)$. This yields
\begin{align}
    A&=L \log \left(-\frac{\sqrt{\cosh(\eta_c)^2-\cosh(\eta_\star)^2}+\sinh (\eta_c)}{\sqrt{\cosh(\eta_c)^2-\cosh(\eta_\star)^2}-\sinh (\eta_c)}\right)\nonumber\\&=L \log \left(\frac{R+\sqrt{L^2+R^2-L^2 \,\coth(T_0/2)^2}}{R-\sqrt{L^2+R^2-L^2\,\coth(T_0/2)^2}}\right),
\end{align}
where we introduced $R=L\sinh(\eta_c)$.
For the sake of completeness, the entanglement entropy follows from the area:
\begin{align}
S_\text{EE}=\frac{1}{6} c \log \left(\frac{R+\sqrt{\frac{c \mu }{12 \pi -12 \pi  \cosh (T_0)}+R^2}}{R-\sqrt{\frac{c \mu }{12 \pi -12 \pi  \cosh (T_0)}+R^2}}\right),\label{eq:EEcomplete}
\end{align}
since $L=\sqrt{c\mu/(24\pi)}$ and $c=3L/(2 G_N)$. 
The expression for the entanglement entropy \eqref{eq:EEcomplete} has a beautiful symmetry. Introducing $\delta T_0=(\cosh(T_0)-1)$, we can re-write eq. \eqref{eq:EEcomplete} as
\begin{equation}
S_\text{EE}=\frac{c}{6}  \log \!\left(\frac{24\pi R_\text{eff}^2}{c \mu }\! \left(\sqrt{1-\frac{c \mu }{12\pi R_\text{eff}^2}}+1\right)\!-1\!\right)\!,\label{eq:EEsym}
\end{equation}
where we also introduced $R^2_\text{eff}=\delta T_0 \,R^2$. We can compensate a larger Euclidean  time interval $T_0$ by making $R$ smaller, i.e. moving the cutoff surface into the bulk. This symmetry is similar to what we discovered in our previous work \cite{Grieninger:2019zts}.

The condition $0<\eta_\star<\eta_c$ imposes $0<2\arccoth(e^{T_0/2})<\eta_c$ or in other words $T_0>\color{black}2\log \left(\coth \left(\frac{\eta_c}{2}\right)\right)=2 \log((\sqrt{L^2+R^2}+L)/R)$. This means that making $R$ smaller puts a lower limit on the timescale we can resolve. Moreover, in the limit $R\to\infty$ (i.e. sending $\eta_c\to \infty)$ the interval behaves as $T_0>\frac{2 L}{R}+\mathcal O(1/R^2)$. Using $L=\sqrt{c\mu/(24\pi)}$, we find $T_0 R>\sqrt{c\mu/(6\pi)}+\mathcal O(1/R^2)$.

For the sake of completeness, we repeated the calculation of this section in a fully radial slicing with metric $\dd s^2=\dd \eta^2+ L^2\sinh(\eta)^2(\dd \tau^2+\dd\phi^2)$ in appendix \ref{app:differentslicing}. This leads to similar results. 

As a final remark, the expression for undeformed AdS$_3$ (no $T\bar T$ deformation) simply follows by expanding eq. \eqref{eq:EEsym} for large $R=L\,\sinh(\eta_c)$:
\begin{equation}
S_\text{EE}= \frac{c}{3}  \log\! \left(\!\frac{\sqrt{\delta T_0} R}{L}\right)\! =\frac{c}{3}  \log \!\left(\frac{ R_\text{eff}}{L }\right)\!,\label{eq:EEsym2}
\end{equation}
where we re-instated $L=\sqrt{c\mu/(24\pi)}$. Note that $\delta T_0=(\cosh(T_0)-1)$ and thus $\sqrt{T_0}=\sqrt{\cosh(T_0)-1}=\sqrt{2}\sinh(T_0/2)$. Remarkably, this looks like the entanglement entropy of a CFT at finite temperature. This confirms our conjecture that integrating over an interval in Euclidean time corresponds to a coarse-grained theory, which in the conformal limit corresponds to a finite-temperature CFT.

\section{Finite temperature}
Let us now briefly discuss the case of finite temperature, i.e. compactified Euclidean time direction. 
There are two cases we can consider \cite{Witten:1998zw}. On the one hand, we can realize thermal effects by thermal AdS, which is the global AdS spacetime of the previous section with compactified time direction i.e.
\begin{equation}
    \dd s^2=\left(1+\frac{\eta^2}{L^2}\right)\dd \tau^2+\left(1+\frac{\eta^2}{L^2}\right)^{-1}\dd\eta^2+\eta^2\dd\Omega^2_{d-1}.
\end{equation}
In this case, the temperature is given by the inverse of the compactified Euclidean time direction.

On the other hand, we can also consider a second spacetime: the asymptotically AdS Schwarzschild black hole described by the Euclidean metric
\begin{equation}\label{eq:schwarz}
    \dd s^2=f(\eta)\dd\tau^2+\frac{\dd\eta^2}{f(\eta)}+\eta^2\dd\Omega^2_{d-1},\ f(\eta)=1-\frac{\mu}{\eta^{d-2}}+\frac{\eta^2}{L^2}.
\end{equation}
Note that in $d=2$, the metric of the AdS$_3$ Schwarzschild black hole matches the metric of the non-rotating BTZ black hole \cite{Banados:1992wn} with $\mu-1\sim M$.
The temperature of the AdS Schwarzschild black hole is given by
\begin{equation}
    T=\frac{(d-2) L^2+d \eta_h^2}{4 \pi  L^2 \eta_h},
\end{equation}
where the horizon location $\eta_h$ is given by the largest root of $f(\eta_h)=0$.
For AdS$_3$, the temperature is $T=\frac{\eta_h}{2L^2\pi}=\frac{\sqrt{-1+\mu}}{2L\pi}$. 

By computing the free energies it turns out that thermal AdS is preferred for $T\leq1/(2\pi L)$ (i.e. $\eta_h\le L)$, while for $T>1/(2\pi L)$, the Schwarzschild black hole is the thermodynamically favored solution, i.e. there is a phase transition at temperature $T=\frac{d-1}{2\pi L}$, the famous Hawking-Page phase transition \cite{Hawking:1982dh}.
As we will see, the temporal entanglement entropy jumps at the phase transition, when we are transitioning from one space-time to the other. Temporal entanglement as an order parameter of the Hagedorn deconfinement transition was considered in \cite{Fujita:2008zv}.

For simplicity we restrict ourselves to $d=2$.  Some related results in the context of \textit{geometric} entropy may be found in \cite{Fujita:2008zv} ($d=4$ case) and \cite{Bah:2008cj}.

In the following, we will do both calculations (Schwarzschild and thermal AdS) at once. We compute the entanglement entropy in the Schwarzschild metric and the results for thermal AdS follow by setting $\mu=0$. 
The minimal surfaces can be derived by considering
\begin{equation}\label{eq:areatherm}
    \mathcal L=\sqrt{\frac{1}{1-\mu+\eta^2/L^2}+(1-\mu+\eta^2/L^2) \tau'(\eta)^2}.
\end{equation}
Since the time $\tau$ does not appear explicitly, we immediately find
\begin{equation}
    \frac{\left( \frac{\eta^2}{L^2}+1-\mu\right) \tau'(\eta)}{\sqrt{\frac{1}{\frac{\eta^2}{L^2}+1-\mu }+\left( \frac{\eta^2}{L^2}+1-\mu\right) \tau'(\eta)^2}}=c_1
\end{equation}
and $c_1=\frac{\sqrt{\eta_\star^2-(\mu -1) L^2}}{L}$ since $\tau'(\eta_\star)=\infty$. Solving the equation of motion for $\tau(\eta)$ yields
\begin{equation}\label{eq:surfacetherm}
   \tau(\eta)=c_2 -\frac{L \arctanh\left(\frac{r}{L \sqrt{-\frac{(\mu -1) (\eta-\eta_\star) (\eta+\eta_\star)}{\eta_\star^2-(\mu -1) L^2}}}\right)}{\sqrt{1-\mu }}, \ c_2=\frac{L\pi}{2\sqrt{\mu-1}},
\end{equation}
where the constant $c_2$ is determined by the condition $\tau(\eta_\star)=0$. We can relate the turning point to the interval traced on the boundary by inverting the relation\newline $\tau(\infty)=-\frac{L }{\sqrt{\mu -1}}\arctan\left(L \sqrt{\frac{\mu -1}{\eta_\star^2-(\mu -1) L^2}}\right)=T_0/2$.
The entanglement entropy is proportional to the area $A$ of the minimal surface which can be computed by integrating \eqref{eq:areatherm} evaluated for the minimal surface \eqref{eq:surfacetherm} from its turning point to the cutoff surface
\begin{equation}
    A=2\int_{\eta_\star}^{\eta_c}\dd\eta\,\mathcal L=2L \arccoth\left(\frac{\eta_\star}{\sqrt{\eta_c^2-\eta_\star^2}}\right).
\end{equation}
Hence, the entanglement entropy is given by
\begin{equation}
    S_\text{EE}=\frac{A}{4G_N}=
    \begin{cases}
    &\frac{L}{2G_N} \arccoth\left(\frac{\eta_c}{\sqrt{\eta_c^2-(\mu -1) L^2 \csc ^2\left(\frac{\sqrt{\mu -1} T_0}{2 L}\right)}}\right) \text{for $\mu>1$},\\
    &\frac{L}{2G_N} \arccoth\left(\frac{\eta_c}{\sqrt{\eta_c^2-L^2 \csch^2\left(\frac{T_0}{2 L}\right)}}\right) \text{for $\mu=0$}.
    \end{cases}
\end{equation}
Similarly to the last section, we observe a symmetry in the cutoff position $\eta_c=R$ and the size of the interval $T_0$ and could introduce an effective radius $R_\text{eff}$. However, in the finite temperature case the symmetry is triple instead of a pair. We can write the $\mu>1$ case of the entanglement entropy in eq. \eqref{eq:timelikeb} as 
\begin{equation}
    S_\text{EE}=\frac{L}{2G_N} \arccoth\left(\frac{1}{\sqrt{1- 4T^2\pi^2 L^4 \csc ^2\left(\pi T T_0\right)/R^2}}\right)=\frac{L}{2G_N} \arccoth\left(\frac{R_\text{eff}}{\sqrt{R_\text{eff}^2-1}}\right),
\end{equation}
where $R_\text{eff}=\eta_c\,\sin(\pi TT_0)/(2\pi T L^2)$. This time we can compensate increasing $T_0$ either by lowering the temperature or decreasing cutoff scale (i.e. integrating out more UV degrees of freedom).

\begin{figure}
    \centering
    \includegraphics[width=0.6\linewidth]{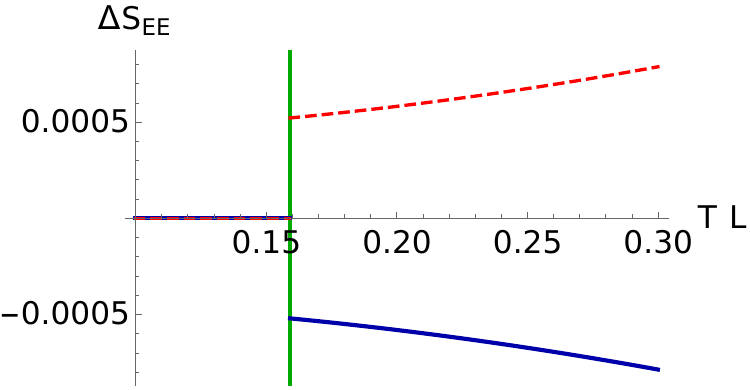}
    \caption{Difference in entanglement entropy $\Delta S_\text{EE}\equiv S_\text{EE}(\mu)-S_\text{EE}(\mu=0)$ for a temporal interval $T_0=L/10$ as given by \eqref{eq:timelikeb} (blue) and a spatial interval $X_0=L/10$ as given by \eqref{eq:spacelikeb} (red, dashed). The phase transition from thermal AdS to BTZ is indicated by the vertical green line. For simplicity, we set $G_N=L/2$ and $\eta_c=10^7 L$.}
    \label{fig:PT}
\end{figure}

At large cutoff, $\varepsilon=1/\eta_c\to0$, we can extract the standard form of the entanglement entropy to leading order in $\varepsilon$
\begin{equation}\label{eq:timelikeb}
    S_\text{EE}=\frac{A}{4G_N}=
    \begin{cases}
    &\frac{L}{2G_N} \log \left(\frac{\eta_c \sin (\pi  T T_0)}{\pi  L^2 T}\right) \text{for $\mu>1$},\\
    &\frac{L}{2G_N} \log \left(\frac{2 \eta_c \sinh \left(\frac{T_0}{2 L}\right)}{L}\right) \text{for $\mu=0$}.
    \end{cases}
\end{equation}
An analogous calculation for the entanglement entropy of a spatial cut $X_0$ yields (see appendix \ref{app:sl})
\begin{equation}\label{eq:spacelikeb}
    S_\text{EE}=\frac{A}{4G_N}=
    \begin{cases}
    &\frac{L}{2G_N} \log \left(\frac{\eta_c \sinh (\pi  T X_0)}{\pi  L^2 T}\right) \text{for $\mu>1$},\\
    &\frac{L}{2G_N} \log \left(\frac{2 \eta_c \sin \left(\frac{X_0}{2 L}\right)}{L}\right) \text{for $\mu=0$}.
    \end{cases}
\end{equation}
We visualize the result in eq. \eqref{eq:timelikeb} and eq. \eqref{eq:spacelikeb} in figure \ref{fig:PT}. In both cases, we observe a first order jump at the phase transition from thermal AdS to the BTZ geometry.

Note that setting $\eta_c\to \frac{\eta_c}{2 \pi  L T}+\mathcal O(1/\eta_c)$ and $X_0\to 2\pi T T_0$ and $T\to (2L\pi)^{-1}$ maps the finite temperature case for $T_0$ ($X_0$) to the thermal AdS case for the $X_0$ ($T_0$). Recall that for $\mu=0$, the periodicity is $\tau\to\tau+2\pi L$ ($x\to x+2\pi L$). For more details about this symmetry see appendix \ref{app:relbtztherm}.

There is a second type of minimal surface that consists of three pieces. At $\tau=T_0/2$ it reaches from the boundary to the horizon, then extends along the horizon $\eta=\eta_h$ and goes back to the boundary at $\tau=-T_0/2$.
The area of this surface may be computed by summing up the three contributions, i.e.
\begin{equation}
    A_2=2\int_{\eta_\star}^{\eta_\infty}\dd\eta\sqrt{\frac{1}{1-\mu+r^2/L^2}}=2L\log \left(\frac{\sqrt{\eta_c^2-4 \pi ^2 L^4 T^2}+\eta_c}{2\pi\,L^2\,T}\right),
\end{equation}
which is independent of $T_0$. Note that the contribution along the horizon vanishes. To find out which surface is preferred, we examine if their difference becomes positive
\begin{equation}
    A-A_2>0, 
\end{equation}
which is never the case. This difference approaches zero when the sine reaches its peak i.e. for $T_0=1/(2T)$.

\section{The nonrelativistic case}
To conclude our discussion, we work out a simple example where temporal entanglement entropy has access to information that cannot be obtained from spatial entanglement.
In the following, we consider simple non-relativistic geometries at zero temperature described by the metric~\cite{Charmousis:2010zz,Ogawa:2011bz,Huijse:2011ef,Dong:2012se}
\begin{equation}
    \dd s^2=\eta^{2\frac{d_e-1}{d-1}}\left(\eta^{2z-2}\dd\tau^2+\frac{\dd \eta^2}{\eta^4}+\dd\bm{x}^2\right), \ d_e=d-\theta,
\end{equation}
where $z$ is the dynamical critical exponent and $\theta$ the hyperscaling violation exponent. The metric reduces to AdS$_{d+1}$ if we set $(d_e,z)=(d,1)$. We can find the extremal surfaces by minimizing
\begin{equation}
    \mathcal L=\eta ^{d_e-3} \sqrt{\eta ^{2 z+2} \tau'(\eta )^2+1},
\end{equation}
which is notably independent of the number of spatial dimensions. Since the Lagrangian does not explicitly depend on time, we immediately find
\begin{equation}
    \frac{\eta ^{d_e+2 z-1} \tau'(\eta )}{\sqrt{\eta ^{2 z+2} \tau'(\eta )^2+1}}=c_1.
\end{equation}
The derivative $\tau'(\eta)$ diverges at the turning point $\tau'(\eta_\star)=\infty$ and the constant is determined by $c_1=(\eta_\star)^{d_e+z-2}$.
Performing the $\eta$ integration, the minimal surface is given by
\begin{align*}
    \tau(\eta)=c_2&-\frac{\eta ^{-2 d_e-4 z+2} \eta_\star^{d_e+z-2} \sqrt{\eta ^{2 (d_e+2 z)}-\eta ^{2 z+4} \eta_\star^{2 (d_e+z-2)}}}{d_e+2 z-2} \\&\times\ _2F_1\left(1,\frac{2 d_e+3 z-4}{2 (d_e+z-2)};2-\frac{d_e-2}{2 (d_e+z-2)};\left(\frac{\eta }{\eta_\star}\right)^{-2 (d_e+z-2)}\right).
\end{align*}
The physically sensible ranges for the parameters $(d_e,z)$ were worked out in \cite{Dong:2012se} using the null energy condition and positivity of the specific heat (for a summary of the ranges see eq. (1.8) in \cite{Jeong:2022jmp}. Note that $d$ is shifted by one since we consider the bulk to be $d+1$ dimensional). We will pay special attention to the case $\theta=d-2$ (i.e. $d_e=2$) where \cite{Dong:2012se} showed that the spatial entanglement entropy gives a logarithmic instead of a power-law dependence. The authors argued that in this case there is a strongly coupled Fermi surface appearing in the dual field theory (for more details see \cite{Dong:2012se,Ogawa:2011bz,Huijse:2011ef}).
Hence, we restrict ourselves to the case $d_e\geq 2$ and $z\geq1$ in the following.
 In this case, the constant $c_2$ which is determined by the condition $t(\eta_\star)=0$ reads
\begin{equation}
    c_2=\frac{\sqrt{\pi } \eta_\star^{-z} \Gamma \left(2-\frac{d_e-2}{2 (d_e+z-2)}\right)}{(d_e+2 z-2) \Gamma \left(\frac{2 d_e+3 z-4}{2 (d_e+z-2)}\right)}.
\end{equation}
With the full expression for the geodesics at hand, we can relate the turning point to the interval traced at the boundary
\begin{equation}
    \eta_\star= 2^{1/z} \pi ^{\frac{1}{2 z}} \left(\frac{T_0 z \Gamma \left(\frac{z}{2 (d_e+z-2)}\right)}{\Gamma \left(\frac{d_e+2 z-2}{2 (d_e+z-2)}\right)}\right)^{-1/z}.
\end{equation}
The authors of \cite{Dong:2012se,Khoeini-Moghaddam:2020ymm,Paul:2020gou,Jeong:2022jmp} showed that the regularized spatial entanglement in the case $d_e=2$ is proportional to the logarithm of the considered interval (instead of a power law).
Setting $d_e$=2 and plugging our solution back into the area functional, we find that the area is given by
\begin{equation}
    A=2\int_{\eta_\star}^{\eta_c}\!\dd\eta\,  \mathcal L=\frac{2}{z}\arctanh\!\left(\!\sqrt{1-\left(\frac{\eta_\star}{\eta_c}\right)^{2z}}\right)=\frac{2}{z}\arctanh\!\left(\!\sqrt{1-4\left(\eta_c^{2z}T_0^2z^2\right)^{-1}}\right).
\end{equation}
We can connect to the usual (undeformed) case by sending the cutoff surface to the boundary $\eta_c\to\infty$ and examine the leading contributions in $\varepsilon=1/\eta_c$ which are given by
\begin{align}
    S_\text{EE}=\frac{1}{2\,G_N}\left(\log\left(z^z\,\frac{T_0^{1/z}}{\varepsilon}\right)-\frac{\varepsilon^{2z}}{z^z\,T_0^2}+\ldots\right).
\end{align}
We note that while the spatial entanglement entropy is insensitive to the dynamical exponent $z$, we can extract information about it by examining the scaling of the entanglement entropy with the size of an Euclidean time interval.

If we consider $T\bar T$ deformed theories, we can keep the cutoff finite. The extensions of the $T\bar T$/cutoff AdS conjecture for hyperscaling violating geometries were discussed in \cite{Alishahiha:2019lng,Jeong:2022jmp,Khoeini-Moghaddam:2020ymm}. Similarly to the previous section, there is a restriction on the smallest time interval (defined in terms of the undeformed theory) that we can resolve. In the case of this subsection, i.e. $d_e=2$ and $z>0$, we find the requirement
$z\,T_0\,\eta_c^z>2$ (by imposing $\eta_\star<\eta_c)$. The entanglement entropy for the smallest time interval that we can resolve tends to zero. By introducing $\delta=T_0-\frac{2}{z\,\eta_c^z}$, we can show that the falloff behavior scales to lowest order in $\delta\to 0$ as
\begin{equation}
    S_\text{EE}=\frac{1}{2 G_N} \sqrt{\frac{\eta_c^z}{z}\,\delta}.
\end{equation}

\section{Conclusions}
In conclusion, our exploration of temporal  entanglement entropy reveals intriguing connections between quantum information theory and the renormalization group (RG) flow in quantum field theory (QFT). 
Our proposal associates tracing over Euclidean time with the concept of coarse-graining in QFT. It links the temporal entanglement entropy to  momentum space entanglement. Intriguingly, the temporal entanglement entropy is real-valued unlike the timelike EE in Minkowski space which is complex. 

The authors of \cite{Balasubramanian:2011wt} laid out the concept of momentum space entanglement between different field modes. To our knowledge, there is no universal recipe for computing this momentum space entanglement entropy (which contains information about RG flow) so far. In this manuscript, we propose a clear recipe for how to compute momentum space entanglement entropy using temporal EE (tEE): Wick-rotate the theory to Euclidean time, then consider an interval in Euclidean time and compute the corresponding tEE  which yields the momentum space entanglement entropy. The Euclidean scale $\tau_0$ is inversely proportional to the cutoff in momentum space.

We employ holography, a framework intrinsically linked to (holographic) RG flow, to exemplify our findings. Using the irrelevant $T\bar T$ deformation, we show that there is a one-to-one relation between the UV cutoff and the shortest time interval that can be resolved within the effective (deformed) theory. Integrating out more UV modes corresponds to increasing the minimal $\Delta T_0$ that can be resolved and vice versa. Remarkably, the entanglement entropy for a Euclidean time interval looks like the entanglement entropy of a CFT at finite temperature. Moreover, on the level of the entanglement entropy we observe the following symmetry: making the time interval $T_0$ larger can be compensated by moving the cutoff surface further into the bulk, i.e. integrating out more UV degrees of freedom. By symmetry, we mean that the entanglement entropies are formally equivalent. This symmetry is similar to what we discovered in our previous work \cite{Grieninger:2019zts}. It would be interesting to relate our findings to \cite{Casini:2012ei,Casini:2023kyj}. Introducing finite temperature, we find that the symmetry is enhanced: we have an interplay of temperature, size of the timelike interval and UV cutoff that can all be rotated into one another. For example, increasing the time interval can be compensated by either decreasing the temperature or integrating out more UV degrees of freedom (moving the cutoff inwards). 
Moreover, in the finite temperature case we find that the temporal entanglement entropy for the BTZ black hole can be mapped onto the spatial entanglement entropy in thermal AdS and vice versa by performing the well known $SL(2, Z)$ transformation.

Finally, we show that tEE can by used to detect the dynamical critical exponent $z$ in non-relativistic Lifshitz geometries. By measuring the dependence of the tEE on the size of the Euclidean time interval we can compute the critical exponent which cannot be accessed (at zero temperature and density) through the conventional spacelike entanglement entropy in the static patch. It would be interesting to generalize our discussion of  non-relativistic geometries to finite temperature. Moreover, it would be interesting to compare spatial and temporal entanglement entropies at finite hyperscaling violating exponent $\theta$ and finite critical exponent $z$. We leave these tasks for future work.
\newline\newline
\noindent\textbf{Acknowledgments:} We thank Hyun-Sik Jeong, Andreas Karch, Rene Meyer, Mark Mezei, Tadashi Takayanagi, Yusuke Taki for helpful discussions.
This work was supported by the U.S. Department of Energy under Grants DE-FG88ER4038 (SG, DK) and DE-SC0012704 (DK), and the U.S. Department of Energy, Office of Science, National Quantum Information Science Research Centers, Co-design Center for Quantum Advantage (C2QA) under Contract No.DE-SC0012704 (KI, DK).
\appendix
\section{Different radial slicing}\label{app:differentslicing}
Consider the radial slicing: 
\begin{equation}
    \dd s^2=\dd \eta^2+L^2\sinh(\eta)^2\,(\dd \tau^2+\dd\phi^2).
\end{equation}
The minimal surfaces read
\begin{equation}
 \tau=   \arctan\left(\frac{\sqrt{2} \cosh (\eta) \sinh (\eta_\star)}{\sqrt{\cosh (2 \eta)-\cosh (2 \eta_\star)}}\right)-\frac{\pi }{2}.
\end{equation}
As pointed out in \cite{Grieninger:2019zts} these entangling surfaces are the family of entangling surfaces that we found in \cite{Geng:2019bnn} in the context of the DS/dS correspondence \cite{Alishahiha:2004md} but for AdS$_{2}$ sliced AdS$_3$.
Note that time here seems to be restricted to $\tau\in[-\pi,\pi]$.
The associated area is given by
\begin{equation}
    A=\log\!\left(\sech(\eta_\star)\left(\cosh (\eta_c) + \sqrt{\cosh ( \eta_c)^2-\cosh ( \eta_\star)^2}\right)\right).
\end{equation}
The turning point is related to the length of the time interval at the boundary by $T_0=\tau_+(\infty)-\tau_-(\infty)=2 \arctan(\sinh (\eta_\star))$. Requiring $0<\eta_\star<\eta_c$ implies $0<\arcsinh(\cot(T_0/2))<\eta_c$ or $T_0>24 \arccot\left(e^{\eta c}\right)\in[0,2\pi]$.

\section{Relationship between BTZ and thermal AdS}\label{app:relbtztherm}
This review about the relationship between the BTZ black hole and thermal AdS is taken from \cite{Hubeny:2009rc}.
In Lorentzian signature, the rotation BTZ spacetime is captured by the metric
\begin{equation}\label{eq:rotBTZ}
    \dd s^2=-\frac{(\eta^2-\eta_+^2)(\eta^2-\eta_-^2)}{L^2\eta^2}\dd t^2+\frac{\eta^2 L^2}{(\eta^2-\eta_+^2)(\eta^2-\eta_-^2)}\dd \eta^2+\eta^2\left(\dd \phi-\frac{\eta_+\eta_-}{L\eta^2}\dd t\right)^2,
\end{equation}
with mass $M=(\eta_+^2+\eta_-^2)/L^2$ and angular momentum $J=2/L\,\eta_+\eta_-$. The metric is periodic under 
\begin{equation}
    (t,\phi)\to(t+\frac{i}{T_\text{BTZ},\,\phi+\frac{i\Omega}{T_\text{BTZ}}}),
\end{equation}
with \begin{equation}
    T_\text{BTZ}=\frac{\eta_+^2-\eta_-^2}{2\pi \eta_+ L^2}, \Omega=\frac{\eta_-}{\eta_+L},
\end{equation}
as well as $(t,\phi)\to(t,\phi+2\pi)$.
Performing the coordinate transformation 
\begin{equation}
    \tilde t=-i\left(\frac{\eta_-}{L}t-\eta_+\phi\right),\ \tilde \phi=-i\left(\frac{\eta_+}{L^2}t-\frac{\eta_-}{L}\phi\right), \,\tilde \eta=L\sqrt{\frac{\eta^2-\eta_+^2}{\eta_+^2-\eta_-^2}}
\end{equation}
maps the metric \eqref{eq:rotBTZ} to global AdS$_3$
\begin{equation}
    \dd s^2=-\left(\frac{\tilde \eta^2}{L^2}+1\right)^2\dd\tilde t^2+\frac{\dd \tilde \eta^2}{\frac{\tilde \eta^2}{L^2}+1}+\tilde \eta^2\dd\tilde\phi^2,
\end{equation}
which is periodic under
\begin{equation}
    (\tilde t,\tilde \phi)\to(\tilde t,\tilde \phi+2\pi)\text{    and    } (\tilde t,\tilde \phi)\to(\tilde t+\frac{i}{T_\text{AdS}},\tilde \phi+\frac{i\Omega}{T_\text{AdS}}), \text{    with    } T_\text{AdS}=\frac{1-\Omega^2L^2}{4\pi^2 T_\text{BTZ}L^2}.
\end{equation}
In our case we utilize the non-rotating solution $\eta_-=0$ which implies that $\Omega=0$ and the coordinate transformation simplifies to
\begin{equation}
    \tilde \tau=i\tilde t=-\eta_+\phi,\ \tilde \phi=-i\frac{\eta_+}{L^2}t=\frac{\eta_+}{L^2}\tau, \,\tilde \eta=L\sqrt{\frac{\eta^2}{\eta_+^2}-1}.
\end{equation}
\section{Entanglement Entropy for spacelike cut}\label{app:sl}
In this section we briefly outline the calculation of the entanglement entropy for a spacelike cut in the background \eqref{eq:schwarz} (see for example \cite{Bah:2007kcs,Bah:2008cj}).
For $d=2$, the extremal surface follows from 
\begin{equation}
   \mathcal L= \sqrt{\frac{1}{-\mu +\frac{\eta^2}{L^2}+1}+\eta^2 \left(\frac{\partial x(\eta)}{\partial \eta}\right)^2}
\end{equation}
and reads
\begin{equation}
    x(\eta)=-\frac{\arctanh\left(\frac{\sqrt{\mu -1} L \sqrt{\eta_\star^2-\eta^2}}{\eta_\star \sqrt{(\mu -1) L^2-\eta^2}}\right)}{\sqrt{\mu -1}}.
\end{equation}
The corresponding entanglement entropy is given by 
\begin{align}
    S_\text{EE}&=\frac{L}{2G_N} \arcsinh\left(\sqrt{\frac{(\eta_c-\eta_\star) (\eta_c+\eta_\star)}{\eta_\star^2-(\mu -1) L^2}}\right)\\&=\frac{L}{2G_N}  \arcsinh\left(\sqrt{\frac{\eta_c^2 \sinh ^2\left(\frac{1}{2} \sqrt{\mu -1}\, X_0\right)}{(\mu -1) L^2}-\cosh ^2\left(\frac{1}{2} \sqrt{\mu -1}\, X_0\right)}\right),
\end{align}
where we used that $\eta_\star= -L\sqrt{\mu -1}\,\coth \left(\frac{1}{2} \sqrt{\mu -1} X_0\,\right)$. 

At large cutoff, $\varepsilon=1/\eta_c\to0$, we can extract the standard expression to leading order in $\varepsilon$
\begin{equation}
    S_\text{EE}=\frac{L}{2G_N} \log \left(\frac{\sinh (\pi  L T \,X_0)}{\pi  L^2 T\,\varepsilon}\right),
\end{equation}
where we used that $T=\frac{\sqrt{\mu -1}}{2 \pi  L}$.
\bibliographystyle{JHEP}
\bibliography{references}

\providecommand{\href}[2]{#2}\begingroup\raggedright\begin{thebibliography}{10}

\bibitem{Doi:2023zaf}
K.~Doi, J.~Harper, A.~Mollabashi, T.~Takayanagi and Y.~Taki, \emph{{Timelike
  entanglement entropy}},
  \href{https://doi.org/10.1007/JHEP05(2023)052}{\emph{JHEP} {\bfseries 05}
  (2023) 052} [\href{https://arxiv.org/abs/2302.11695}{{\ttfamily
  2302.11695}}].

\bibitem{Doi:2022iyj}
K.~Doi, J.~Harper, A.~Mollabashi, T.~Takayanagi and Y.~Taki,
  \emph{{Pseudoentropy in dS/CFT and Timelike Entanglement Entropy}},
  \href{https://doi.org/10.1103/PhysRevLett.130.031601}{\emph{Phys. Rev. Lett.}
  {\bfseries 130} (2023) 031601}
  [\href{https://arxiv.org/abs/2210.09457}{{\ttfamily 2210.09457}}].

\bibitem{Reddy:2022zgu}
K.S.~Reddy, \emph{{A timelike entangled island at the initial singularity in a
  JT FLRW ($\Lambda>0$) universe}},
  \href{https://arxiv.org/abs/2211.14893}{{\ttfamily 2211.14893}}.

\bibitem{Diaz:2021snw}
N.L.~Diaz, J.M.~Matera and R.~Rossignoli, \emph{{Path Integrals from Spacetime
  Quantum Actions}},  \href{https://arxiv.org/abs/2111.05383}{{\ttfamily
  2111.05383}}.

\bibitem{Li:2022tsv}
Z.~Li, Z.-Q.~Xiao and R.-Q.~Yang, \emph{{On holographic time-like entanglement
  entropy}}, \href{https://doi.org/10.1007/JHEP04(2023)004}{\emph{JHEP}
  {\bfseries 04} (2023) 004}
  [\href{https://arxiv.org/abs/2211.14883}{{\ttfamily 2211.14883}}].

\bibitem{Liu:2022ugc}
B.~Liu, H.~Chen and B.~Lian, \emph{{Entanglement Entropy of Free Fermions in
  Timelike Slices}},  \href{https://arxiv.org/abs/2210.03134}{{\ttfamily
  2210.03134}}.

\bibitem{Narayan:2022afv}
K.~Narayan, \emph{{de Sitter space, extremal surfaces, and time entanglement}},
  \href{https://doi.org/10.1103/PhysRevD.107.126004}{\emph{Phys. Rev. D}
  {\bfseries 107} (2023) 126004}
  [\href{https://arxiv.org/abs/2210.12963}{{\ttfamily 2210.12963}}].

\bibitem{Alshal:2023kcd}
H.~Alshal, \emph{{Einstein\textquoteright{}s equations and the pseudo-entropy
  of pseudo-Riemannian information manifolds}},
  \href{https://doi.org/10.1007/s10714-023-03130-7}{\emph{Gen. Rel. Grav.}
  {\bfseries 55} (2023) 86} [\href{https://arxiv.org/abs/2301.13017}{{\ttfamily
  2301.13017}}].

\bibitem{Foligno:2023dih}
A.~Foligno, T.~Zhou and B.~Bertini, \emph{{Temporal Entanglement in Chaotic
  Quantum Circuits}},
  \href{https://doi.org/10.1103/PhysRevX.13.041008}{\emph{Phys. Rev. X}
  {\bfseries 13} (2023) 041008}
  [\href{https://arxiv.org/abs/2302.08502}{{\ttfamily 2302.08502}}].

\bibitem{Chen:2023prz}
H.-Y.~Chen, Y.~Hikida, Y.~Taki and T.~Uetoko, \emph{{Complex saddles of
  three-dimensional de Sitter gravity via holography}},
  \href{https://doi.org/10.1103/PhysRevD.107.L101902}{\emph{Phys. Rev. D}
  {\bfseries 107} (2023) L101902}
  [\href{https://arxiv.org/abs/2302.09219}{{\ttfamily 2302.09219}}].

\bibitem{Chen:2023gnh}
Z.~Chen, \emph{{Complex-valued Holographic Pseudo Entropy via Real-time AdS/CFT
  Correspondence}},  \href{https://arxiv.org/abs/2302.14303}{{\ttfamily
  2302.14303}}.

\bibitem{Jiang:2023ffu}
X.~Jiang, P.~Wang, H.~Wu and H.~Yang, \emph{{Timelike entanglement entropy and
  $T\bar{T}$ deformation}},  \href{https://arxiv.org/abs/2302.13872}{{\ttfamily
  2302.13872}}.

\bibitem{Narayan:2023ebn}
K.~Narayan and H.K.~Saini, \emph{{Notes on time entanglement and
  pseudo-entropy}},  \href{https://arxiv.org/abs/2303.01307}{{\ttfamily
  2303.01307}}.

\bibitem{Jiang:2023loq}
X.~Jiang, P.~Wang, H.~Wu and H.~Yang, \emph{{Timelike entanglement entropy in
  dS$_{3}$/CFT$_{2}$}},
  \href{https://doi.org/10.1007/JHEP08(2023)216}{\emph{JHEP} {\bfseries 08}
  (2023) 216} [\href{https://arxiv.org/abs/2304.10376}{{\ttfamily
  2304.10376}}].

\bibitem{Chu:2023zah}
C.-S.~Chu and H.~Parihar, \emph{{Time-like entanglement entropy in AdS/BCFT}},
  \href{https://doi.org/10.1007/JHEP06(2023)173}{\emph{JHEP} {\bfseries 06}
  (2023) 173} [\href{https://arxiv.org/abs/2304.10907}{{\ttfamily
  2304.10907}}].

\bibitem{He:2023wko}
S.~He, J.~Yang, Y.-X.~Zhang and Z.-X.~Zhao, \emph{{Pseudo entropy of primary
  operators in $ T\overline{T}/J\overline{T} $-deformed CFTs}},
  \href{https://doi.org/10.1007/JHEP09(2023)025}{\emph{JHEP} {\bfseries 09}
  (2023) 025} [\href{https://arxiv.org/abs/2305.10984}{{\ttfamily
  2305.10984}}].

\bibitem{Franken:2023pni}
V.~Franken, H.~Partouche, F.~Rondeau and N.~Toumbas, \emph{{Bridging the static
  patches: de Sitter holography and entanglement}},
  \href{https://doi.org/10.1007/JHEP08(2023)074}{\emph{JHEP} {\bfseries 08}
  (2023) 074} [\href{https://arxiv.org/abs/2305.12861}{{\ttfamily
  2305.12861}}].

\bibitem{Chen:2023sry}
H.-Y.~Chen, Y.~Hikida, Y.~Taki and T.~Uetoko, \emph{{Complex saddles of
  Chern-Simons gravity and dS3/CFT2 correspondence}},
  \href{https://doi.org/10.1103/PhysRevD.108.066005}{\emph{Phys. Rev. D}
  {\bfseries 108} (2023) 066005}
  [\href{https://arxiv.org/abs/2306.03330}{{\ttfamily 2306.03330}}].

\bibitem{Kawamoto:2023nki}
T.~Kawamoto, S.-M.~Ruan, Y.-k.~Suzuki and T.~Takayanagi, \emph{{A Half de
  Sitter Holography}},  \href{https://arxiv.org/abs/2306.07575}{{\ttfamily
  2306.07575}}.

\bibitem{Chen:2023eic}
D.~Chen, X.~Jiang and H.~Yang, \emph{{Holographic $T\bar{T}$ deformed
  entanglement entropy in dS$_3$/CFT$_2$}},
  \href{https://arxiv.org/abs/2307.04673}{{\ttfamily 2307.04673}}.

\bibitem{Parzygnat:2023avh}
A.J.~Parzygnat, T.~Takayanagi, Y.~Taki and Z.~Wei, \emph{{SVD Entanglement
  Entropy}},  \href{https://arxiv.org/abs/2307.06531}{{\ttfamily 2307.06531}}.

\bibitem{He:2023ubi}
P.-Z.~He and H.-Q.~Zhang, \emph{{Timelike Entanglement Entropy from Rindler
  Method}},  \href{https://arxiv.org/abs/2307.09803}{{\ttfamily 2307.09803}}.

\bibitem{Carignano:2023xbz}
S.~Carignano, C.R.~Marim\'on and L.~Tagliacozzo, \emph{{On temporal entropy and
  the complexity of computing the expectation value of local operators after a
  quench}},  \href{https://arxiv.org/abs/2307.11649}{{\ttfamily 2307.11649}}.

\bibitem{Guo:2023aio}
W.-z.~Guo and J.~Zhang, \emph{{Sum rule for pseudo R\'enyi entropy}},
  \href{https://arxiv.org/abs/2308.05261}{{\ttfamily 2308.05261}}.

\bibitem{Aguilar-Gutierrez:2023tic}
S.E.~Aguilar-Gutierrez, A.K.~Patra and J.F.~Pedraza, \emph{{Entangled universes
  in dS wedge holography}},
  \href{https://doi.org/10.1007/JHEP10(2023)156}{\emph{JHEP} {\bfseries 10}
  (2023) 156} [\href{https://arxiv.org/abs/2308.05666}{{\ttfamily
  2308.05666}}].

\bibitem{Omidi:2023env}
F.~Omidi, \emph{{Pseudo R\'enyi Entanglement Entropies For an Excited State and
  Its Time Evolution in a 2D CFT}},
  \href{https://arxiv.org/abs/2309.04112}{{\ttfamily 2309.04112}}.

\bibitem{Narayan:2023zen}
K.~Narayan, \emph{{Comments on de Sitter space, extremal surfaces and time
  entanglement}},  \href{https://arxiv.org/abs/2310.00320}{{\ttfamily
  2310.00320}}.

\bibitem{Shinmyo:2023eci}
K.~Shinmyo, T.~Takayanagi and K.~Tasuki, \emph{{Pseudo entropy under joining
  local quenches}},  \href{https://arxiv.org/abs/2310.12542}{{\ttfamily
  2310.12542}}.

\bibitem{Guo:2023tjv}
W.-z.~Guo and Y.~Jiang, \emph{{Pseudo entropy and pseudo-Hermiticity in quantum
  field theories}},  \href{https://arxiv.org/abs/2311.01045}{{\ttfamily
  2311.01045}}.

\bibitem{Apolo:2023ckr}
L.~Apolo, P.-X.~Hao, W.-X.~Lai and W.~Song, \emph{{Extremal surfaces in glue-on
  AdS/$T\bar T$ holography}},
  \href{https://arxiv.org/abs/2311.04883}{{\ttfamily 2311.04883}}.

\bibitem{Kanda:2023jyi}
H.~Kanda, T.~Kawamoto, Y.-k.~Suzuki, T.~Takayanagi, K.~Tasuki and Z.~Wei,
  \emph{{Entanglement Phase Transition in Holographic Pseudo Entropy}},
  \href{https://arxiv.org/abs/2311.13201}{{\ttfamily 2311.13201}}.

\bibitem{He:2023syy}
S.~He, Y.-X.~Zhang, L.~Zhao and Z.-X.~Zhao, \emph{{Entanglement and Pseudo
  Entanglement Dynamics versus Fusion in CFT}},
  \href{https://arxiv.org/abs/2312.02679}{{\ttfamily 2312.02679}}.

\bibitem{Balasubramanian:2011wt}
V.~Balasubramanian, M.B.~McDermott and M.~Van~Raamsdonk, \emph{{Momentum-space
  entanglement and renormalization in quantum field theory}},
  \href{https://doi.org/10.1103/PhysRevD.86.045014}{\emph{Phys. Rev. D}
  {\bfseries 86} (2012) 045014}
  [\href{https://arxiv.org/abs/1108.3568}{{\ttfamily 1108.3568}}].

\bibitem{Costa:2022bvs}
M.H.M.~Costa, J.v.d.~Brink, F.S.~Nogueira and G.a.I.~Krein, \emph{{Momentum
  space entanglement from the Wilsonian effective action}},
  \href{https://doi.org/10.1103/PhysRevD.106.065024}{\emph{Phys. Rev. D}
  {\bfseries 106} (2022) 065024}
  [\href{https://arxiv.org/abs/2207.12103}{{\ttfamily 2207.12103}}].

\bibitem{deBoer:1999tgo}
J.~de~Boer, E.P.~Verlinde and H.L.~Verlinde, \emph{{On the holographic
  renormalization group}},
  \href{https://doi.org/10.1088/1126-6708/2000/08/003}{\emph{JHEP} {\bfseries
  08} (2000) 003} [\href{https://arxiv.org/abs/hep-th/9912012}{{\ttfamily
  hep-th/9912012}}].

\bibitem{deBoer:2000cz}
J.~de~Boer, \emph{{The Holographic renormalization group}},
  \href{https://doi.org/10.1002/1521-3978(200105)49:4/6<339::AID-PROP339>3.0.CO;2-A}{\emph{Fortsch.
  Phys.} {\bfseries 49} (2001) 339}
  [\href{https://arxiv.org/abs/hep-th/0101026}{{\ttfamily hep-th/0101026}}].

\bibitem{Fukuma:2002sb}
M.~Fukuma, S.~Matsuura and T.~Sakai, \emph{{Holographic renormalization
  group}}, \href{https://doi.org/10.1143/PTP.109.489}{\emph{Prog. Theor. Phys.}
  {\bfseries 109} (2003) 489}
  [\href{https://arxiv.org/abs/hep-th/0212314}{{\ttfamily hep-th/0212314}}].

\bibitem{Skenderis:2002wp}
K.~Skenderis, \emph{{Lecture notes on holographic renormalization}},
  \href{https://doi.org/10.1088/0264-9381/19/22/306}{\emph{Class. Quant. Grav.}
  {\bfseries 19} (2002) 5849}
  [\href{https://arxiv.org/abs/hep-th/0209067}{{\ttfamily hep-th/0209067}}].

\bibitem{Ryu:2006ef}
S.~Ryu and T.~Takayanagi, \emph{{Aspects of Holographic Entanglement Entropy}},
  \href{https://doi.org/10.1088/1126-6708/2006/08/045}{\emph{JHEP} {\bfseries
  08} (2006) 045} [\href{https://arxiv.org/abs/hep-th/0605073}{{\ttfamily
  hep-th/0605073}}].

\bibitem{Ryu:2006bv}
S.~Ryu and T.~Takayanagi, \emph{{Holographic derivation of entanglement entropy
  from AdS/CFT}},
  \href{https://doi.org/10.1103/PhysRevLett.96.181602}{\emph{Phys. Rev. Lett.}
  {\bfseries 96} (2006) 181602}
  [\href{https://arxiv.org/abs/hep-th/0603001}{{\ttfamily hep-th/0603001}}].

\bibitem{Hubeny:2007xt}
V.E.~Hubeny, M.~Rangamani and T.~Takayanagi, \emph{{A Covariant holographic
  entanglement entropy proposal}},
  \href{https://doi.org/10.1088/1126-6708/2007/07/062}{\emph{JHEP} {\bfseries
  07} (2007) 062} [\href{https://arxiv.org/abs/0705.0016}{{\ttfamily
  0705.0016}}].

\bibitem{KaplanJHU}
J.~Kaplan, \emph{{Lectures on AdS/CFT from the Bottom Up}},
  {\emph{\url{https://sites.krieger.jhu.edu/jared-kaplan/files/2016/05/AdSCFTCourseNotesCurrentPublic.pdf}}
  }.

\bibitem{Miyaji:2015fia}
M.~Miyaji, T.~Numasawa, N.~Shiba, T.~Takayanagi and K.~Watanabe,
  \emph{{Continuous Multiscale Entanglement Renormalization Ansatz as
  Holographic Surface-State Correspondence}},
  \href{https://doi.org/10.1103/PhysRevLett.115.171602}{\emph{Phys. Rev. Lett.}
  {\bfseries 115} (2015) 171602}
  [\href{https://arxiv.org/abs/1506.01353}{{\ttfamily 1506.01353}}].

\bibitem{Miyaji:2015yva}
M.~Miyaji and T.~Takayanagi, \emph{{Surface/State Correspondence as a
  Generalized Holography}},
  \href{https://doi.org/10.1093/ptep/ptv089}{\emph{PTEP} {\bfseries 2015}
  (2015) 073B03} [\href{https://arxiv.org/abs/1503.03542}{{\ttfamily
  1503.03542}}].

\bibitem{Caputa:2017urj}
P.~Caputa, N.~Kundu, M.~Miyaji, T.~Takayanagi and K.~Watanabe, \emph{{Anti-de
  Sitter Space from Optimization of Path Integrals in Conformal Field
  Theories}}, \href{https://doi.org/10.1103/PhysRevLett.119.071602}{\emph{Phys.
  Rev. Lett.} {\bfseries 119} (2017) 071602}
  [\href{https://arxiv.org/abs/1703.00456}{{\ttfamily 1703.00456}}].

\bibitem{Goto:2017olq}
K.~Goto and T.~Takayanagi, \emph{{CFT descriptions of bulk local states in the
  AdS black holes}}, \href{https://doi.org/10.1007/JHEP10(2017)153}{\emph{JHEP}
  {\bfseries 10} (2017) 153}
  [\href{https://arxiv.org/abs/1704.00053}{{\ttfamily 1704.00053}}].

\bibitem{Smirnov:2016lqw}
F.A.~Smirnov and A.B.~Zamolodchikov, \emph{{On space of integrable quantum
  field theories}},
  \href{https://doi.org/10.1016/j.nuclphysb.2016.12.014}{\emph{Nucl. Phys. B}
  {\bfseries 915} (2017) 363}
  [\href{https://arxiv.org/abs/1608.05499}{{\ttfamily 1608.05499}}].

\bibitem{Cavaglia:2016oda}
A.~Cavagli\`a, S.~Negro, I.M.~Sz\'ecs\'enyi and R.~Tateo, \emph{{$T
  \bar{T}$-deformed 2D Quantum Field Theories}},
  \href{https://doi.org/10.1007/JHEP10(2016)112}{\emph{JHEP} {\bfseries 10}
  (2016) 112} [\href{https://arxiv.org/abs/1608.05534}{{\ttfamily
  1608.05534}}].

\bibitem{Zamolodchikov:2004ce}
A.B.~Zamolodchikov, \emph{{Expectation value of composite field T anti-T in
  two-dimensional quantum field theory}},
  \href{https://arxiv.org/abs/hep-th/0401146}{{\ttfamily hep-th/0401146}}.

\bibitem{McGough:2016lol}
L.~McGough, M.~Mezei and H.~Verlinde, \emph{{Moving the CFT into the bulk with
  $ T\overline{T} $}},
  \href{https://doi.org/10.1007/JHEP04(2018)010}{\emph{JHEP} {\bfseries 04}
  (2018) 010} [\href{https://arxiv.org/abs/1611.03470}{{\ttfamily
  1611.03470}}].

\bibitem{Guica:2019nzm}
M.~Guica and R.~Monten, \emph{{$T\bar T$ and the mirage of a bulk cutoff}},
  \href{https://doi.org/10.21468/SciPostPhys.10.2.024}{\emph{SciPost Phys.}
  {\bfseries 10} (2021) 024}
  [\href{https://arxiv.org/abs/1906.11251}{{\ttfamily 1906.11251}}].

\bibitem{Jiang:2019epa}
Y.~Jiang, \emph{{A pedagogical review on solvable irrelevant deformations of 2D
  quantum field theory}},
  \href{https://doi.org/10.1088/1572-9494/abe4c9}{\emph{Commun. Theor. Phys.}
  {\bfseries 73} (2021) 057201}
  [\href{https://arxiv.org/abs/1904.13376}{{\ttfamily 1904.13376}}].

\bibitem{GuicaCERN}
M.~Guica, \emph{{T$\bar T$ deformations and holography (Lecture at CERN Winter
  School on Supergravity, Strings and Gauge Theory 2020)}},
  {\emph{\url{https://indico.cern.ch/event/857396/contributions/3706292/attachments/2036750/3410352/ttbar_cern_v1s.pdf}}
  }.

\bibitem{Hartman:2018tkw}
T.~Hartman, J.~Kruthoff, E.~Shaghoulian and A.~Tajdini, \emph{{Holography at
  finite cutoff with a $T^2$ deformation}},
  \href{https://doi.org/10.1007/JHEP03(2019)004}{\emph{JHEP} {\bfseries 03}
  (2019) 004} [\href{https://arxiv.org/abs/1807.11401}{{\ttfamily
  1807.11401}}].

\bibitem{Kraus:2018xrn}
P.~Kraus, J.~Liu and D.~Marolf, \emph{{Cutoff AdS$_{3}$ versus the $
  T\overline{T} $ deformation}},
  \href{https://doi.org/10.1007/JHEP07(2018)027}{\emph{JHEP} {\bfseries 07}
  (2018) 027} [\href{https://arxiv.org/abs/1801.02714}{{\ttfamily
  1801.02714}}].

\bibitem{Gorbenko:2018oov}
V.~Gorbenko, E.~Silverstein and G.~Torroba, \emph{{dS/dS and $T\bar T$}},
  \href{https://arxiv.org/abs/1811.07965}{{\ttfamily 1811.07965}}.

\bibitem{Grieninger:2019zts}
S.~Grieninger, \emph{{Entanglement entropy and $ T\overline{T} $ deformations
  beyond antipodal points from holography}},
  \href{https://doi.org/10.1007/JHEP11(2019)171}{\emph{JHEP} {\bfseries 11}
  (2019) 171} [\href{https://arxiv.org/abs/1908.10372}{{\ttfamily
  1908.10372}}].

\bibitem{Grieninger:2020wsb}
S.~Grieninger, \emph{{Non-equilibrium dynamics in Holography}}, Ph.D. thesis,
  Jena U., 2020.
\newblock \href{https://arxiv.org/abs/2012.10109}{{\ttfamily 2012.10109}}.
\newblock 10.22032/dbt.45425.

\bibitem{Brown:1986nw}
J.D.~Brown and M.~Henneaux, \emph{{Central Charges in the Canonical Realization
  of Asymptotic Symmetries: An Example from Three-Dimensional Gravity}},
  \href{https://doi.org/10.1007/BF01211590}{\emph{Commun. Math. Phys.}
  {\bfseries 104} (1986) 207}.

\bibitem{deHaro:2000vlm}
S.~de~Haro, S.N.~Solodukhin and K.~Skenderis, \emph{{Holographic reconstruction
  of space-time and renormalization in the AdS / CFT correspondence}},
  \href{https://doi.org/10.1007/s002200100381}{\emph{Commun. Math. Phys.}
  {\bfseries 217} (2001) 595}
  [\href{https://arxiv.org/abs/hep-th/0002230}{{\ttfamily hep-th/0002230}}].

\bibitem{Chen:2018eqk}
B.~Chen, L.~Chen and P.-X.~Hao, \emph{{Entanglement entropy in
  $T\overline{T}$-deformed CFT}},
  \href{https://doi.org/10.1103/PhysRevD.98.086025}{\emph{Phys. Rev. D}
  {\bfseries 98} (2018) 086025}
  [\href{https://arxiv.org/abs/1807.08293}{{\ttfamily 1807.08293}}].

\bibitem{Witten:1998zw}
E.~Witten, \emph{{Anti-de Sitter space, thermal phase transition, and
  confinement in gauge theories}},
  \href{https://doi.org/10.4310/ATMP.1998.v2.n3.a3}{\emph{Adv. Theor. Math.
  Phys.} {\bfseries 2} (1998) 505}
  [\href{https://arxiv.org/abs/hep-th/9803131}{{\ttfamily hep-th/9803131}}].

\bibitem{Banados:1992wn}
M.~Banados, C.~Teitelboim and J.~Zanelli, \emph{{The Black hole in
  three-dimensional space-time}},
  \href{https://doi.org/10.1103/PhysRevLett.69.1849}{\emph{Phys. Rev. Lett.}
  {\bfseries 69} (1992) 1849}
  [\href{https://arxiv.org/abs/hep-th/9204099}{{\ttfamily hep-th/9204099}}].

\bibitem{Hawking:1982dh}
S.W.~Hawking and D.N.~Page, \emph{{Thermodynamics of Black Holes in anti-De
  Sitter Space}}, \href{https://doi.org/10.1007/BF01208266}{\emph{Commun. Math.
  Phys.} {\bfseries 87} (1983) 577}.

\bibitem{Fujita:2008zv}
M.~Fujita, T.~Nishioka and T.~Takayanagi, \emph{{Geometric Entropy and
  Hagedorn/Deconfinement Transition}},
  \href{https://doi.org/10.1088/1126-6708/2008/09/016}{\emph{JHEP} {\bfseries
  09} (2008) 016} [\href{https://arxiv.org/abs/0806.3118}{{\ttfamily
  0806.3118}}].

\bibitem{Bah:2008cj}
I.~Bah, L.A.~Pando~Zayas and C.A.~Terrero-Escalante, \emph{{Holographic
  Geometric Entropy at Finite Temperature from Black Holes in Global Anti de
  Sitter Spaces}}, \href{https://doi.org/10.1142/S0217751X12500480}{\emph{Int.
  J. Mod. Phys. A} {\bfseries 27} (2012) 1250048}
  [\href{https://arxiv.org/abs/0809.2912}{{\ttfamily 0809.2912}}].

\bibitem{Charmousis:2010zz}
C.~Charmousis, B.~Gouteraux, B.S.~Kim, E.~Kiritsis and R.~Meyer,
  \emph{{Effective Holographic Theories for low-temperature condensed matter
  systems}}, \href{https://doi.org/10.1007/JHEP11(2010)151}{\emph{JHEP}
  {\bfseries 11} (2010) 151} [\href{https://arxiv.org/abs/1005.4690}{{\ttfamily
  1005.4690}}].

\bibitem{Ogawa:2011bz}
N.~Ogawa, T.~Takayanagi and T.~Ugajin, \emph{{Holographic Fermi Surfaces and
  Entanglement Entropy}},
  \href{https://doi.org/10.1007/JHEP01(2012)125}{\emph{JHEP} {\bfseries 01}
  (2012) 125} [\href{https://arxiv.org/abs/1111.1023}{{\ttfamily 1111.1023}}].

\bibitem{Huijse:2011ef}
L.~Huijse, S.~Sachdev and B.~Swingle, \emph{{Hidden Fermi surfaces in
  compressible states of gauge-gravity duality}},
  \href{https://doi.org/10.1103/PhysRevB.85.035121}{\emph{Phys. Rev. B}
  {\bfseries 85} (2012) 035121}
  [\href{https://arxiv.org/abs/1112.0573}{{\ttfamily 1112.0573}}].

\bibitem{Dong:2012se}
X.~Dong, S.~Harrison, S.~Kachru, G.~Torroba and H.~Wang, \emph{{Aspects of
  holography for theories with hyperscaling violation}},
  \href{https://doi.org/10.1007/JHEP06(2012)041}{\emph{JHEP} {\bfseries 06}
  (2012) 041} [\href{https://arxiv.org/abs/1201.1905}{{\ttfamily 1201.1905}}].

\bibitem{Jeong:2022jmp}
H.-S.~Jeong, W.-B.~Pan, Y.-W.~Sun and Y.-T.~Wang, \emph{{Holographic study of $
  T\overline{T} $ like deformed HV QFTs: holographic entanglement entropy}},
  \href{https://doi.org/10.1007/JHEP02(2023)018}{\emph{JHEP} {\bfseries 02}
  (2023) 018} [\href{https://arxiv.org/abs/2211.00518}{{\ttfamily
  2211.00518}}].

\bibitem{Khoeini-Moghaddam:2020ymm}
S.~Khoeini-Moghaddam, F.~Omidi and C.~Paul, \emph{{Aspects of Hyperscaling
  Violating Geometries at Finite Cutoff}},
  \href{https://doi.org/10.1007/JHEP02(2021)121}{\emph{JHEP} {\bfseries 02}
  (2021) 121} [\href{https://arxiv.org/abs/2011.00305}{{\ttfamily
  2011.00305}}].

\bibitem{Paul:2020gou}
C.~Paul, \emph{{Quantum entanglement measures from Hyperscaling violating
  geometries with finite radial cut off at general d, $\theta$ from the
  emergent global symmetry}},
  \href{https://arxiv.org/abs/2012.01895}{{\ttfamily 2012.01895}}.

\bibitem{Alishahiha:2019lng}
M.~Alishahiha and A.~Faraji~Astaneh, \emph{{Complexity of Hyperscaling
  Violating Theories at Finite Cutoff}},
  \href{https://doi.org/10.1103/PhysRevD.100.086004}{\emph{Phys. Rev. D}
  {\bfseries 100} (2019) 086004}
  [\href{https://arxiv.org/abs/1905.10740}{{\ttfamily 1905.10740}}].

\bibitem{Casini:2012ei}
H.~Casini and M.~Huerta, \emph{{On the RG running of the entanglement entropy
  of a circle}}, \href{https://doi.org/10.1103/PhysRevD.85.125016}{\emph{Phys.
  Rev. D} {\bfseries 85} (2012) 125016}
  [\href{https://arxiv.org/abs/1202.5650}{{\ttfamily 1202.5650}}].

\bibitem{Casini:2023kyj}
H.~Casini, I.~Salazar~Landea and G.~Torroba, \emph{{Irreversibility, QNEC, and
  defects}}, \href{https://doi.org/10.1007/JHEP07(2023)004}{\emph{JHEP}
  {\bfseries 07} (2023) 004}
  [\href{https://arxiv.org/abs/2303.16935}{{\ttfamily 2303.16935}}].

\bibitem{Geng:2019bnn}
H.~Geng, S.~Grieninger and A.~Karch, \emph{{Entropy, Entanglement and Swampland
  Bounds in DS/dS}}, \href{https://doi.org/10.1007/JHEP06(2019)105}{\emph{JHEP}
  {\bfseries 06} (2019) 105}
  [\href{https://arxiv.org/abs/1904.02170}{{\ttfamily 1904.02170}}].

\bibitem{Alishahiha:2004md}
M.~Alishahiha, A.~Karch, E.~Silverstein and D.~Tong, \emph{{The dS/dS
  correspondence}}, \href{https://doi.org/10.1063/1.1848341}{\emph{AIP Conf.
  Proc.} {\bfseries 743} (2004) 393}
  [\href{https://arxiv.org/abs/hep-th/0407125}{{\ttfamily hep-th/0407125}}].

\bibitem{Hubeny:2009rc}
V.E.~Hubeny, D.~Marolf and M.~Rangamani, \emph{{Hawking radiation from AdS
  black holes}},
  \href{https://doi.org/10.1088/0264-9381/27/9/095018}{\emph{Class. Quant.
  Grav.} {\bfseries 27} (2010) 095018}
  [\href{https://arxiv.org/abs/0911.4144}{{\ttfamily 0911.4144}}].

\bibitem{Bah:2007kcs}
I.~Bah, A.~Faraggi, L.A.~Pando~Zayas and C.A.~Terrero-Escalante,
  \emph{{Holographic entanglement entropy and phase transitions at finite
  temperature}}, \href{https://doi.org/10.1142/S0217751X0904542X}{\emph{Int. J.
  Mod. Phys. A} {\bfseries 24} (2009) 2703}
  [\href{https://arxiv.org/abs/0710.5483}{{\ttfamily 0710.5483}}].

\end{thebibliography}\endgroup

\end{document}